\documentclass[letter,longauth]{aa}  
\usepackage{graphicx}
\usepackage{txfonts}
\begin{document}

   \title{Temporal morphological changes in the Imhotep region of comet 67P/Churyumov-Gerasimenko}

   \author{
O.~Groussin\inst{1}
\and H.~Sierks\inst{2}
\and C.~Barbieri\inst{3}
\and P.~Lamy\inst{1}
\and R.~Rodrigo\inst{4,5}
\and D.~Koschny\inst{6}
\and H.~Rickman\inst{7,8}
\and H.~U.~Keller\inst{9,10}
\and M.~F.~A'Hearn\inst{11}
\and A.-T.~Auger\inst{1,12}
\and M.~A.~Barucci\inst{13}
\and J.-L.~Bertaux\inst{14} 
\and I.~Bertini\inst{15} 
\and S.~Besse\inst{6}
\and G.~Cremonese\inst{16} 
\and V.~Da Deppo\inst{17} 
\and B.~Davidsson\inst{7} 
\and S.~Debei\inst{16} 
\and M.~De Cecco\inst{18}
\and M.~R.~El-Maarry\inst{19} 
\and S.~Fornasier\inst{13} 
\and M.~Fulle\inst{20}
\and P.~J.~Guti\'errez\inst{21} 
\and C.~G\"uttler\inst{2} 
\and S.~Hviid\inst{3} 
\and W.-H Ip\inst{22} 
\and L.~Jorda\inst{1} 
\and J.~Knollenberg\inst{3} 
\and G.~Kovacs\inst{2}
\and J.~R.~Kramm\inst{2}
\and E.~K\"uhrt\inst{3} 
\and M.~K\"uppers\inst{23}
\and L.~M.~Lara\inst{21} 
\and M.~Lazzarin\inst{3} 
\and J.~J.~Lopez Moreno\inst{21} 
\and S.~Lowry\inst{24} 
\and S.~Marchi\inst{25} 
\and F.~Marzari\inst{3} 
\and M. Massironi\inst{15,26} 
\and S.~Mottola\inst{3}
\and G.~Naletto\inst{15,17,27} 
\and N.~Oklay\inst{2} 
\and M.~Pajola\inst{15}
\and A.~Pommerol\inst{19} 
\and N.~Thomas\inst{19}
\and I.~Toth\inst{28}
\and C.~Tubiana\inst{2}
\and J.-B.~Vincent\inst{2}
        }
   \institute{\tiny Aix Marseille Universit\'e, CNRS, LAM (Laboratoire d'Astrophysique de Marseille) UMR 7326, 13388, Marseille, France 
         \and
         Max-Planck-Institut f\"ur Sonnensystemforschung, 37077 G\"ottingen, Germany
         \and
         Department of Physics and Astronomy, Padova University, Vicolo dell'Osservatorio 3, 35122, Padova, Italy
         \and
         Centro de Astrobiologia (INTA-CSIC), 28691 Villanueva de la Canada, Madrid, Spain 
         \and
         International Space Science Institute, Hallerstrasse 6, CH-3012 Bern, Switzerland 
         \and
         Scientific Support Office, European Space Agency, 2201, Noordwijk, The Netherlands
         \and
	 Department of Physics and Astronomy, Uppsala University, Box 516, 75120, Uppsala, Sweden
         \and
	 PAS Space Research Center, Bartycka 18A, PL-00716 Warszawa, Poland
         \and
         Institute of Planetary Research, DLR, Rutherfordstrasse 2, 12489, Berlin, Germany
	 \and
         Institute for Geophysics and Extraterrestrial Physics, TU Braunschweig, 38106, Germany
         \and
	 Department of Astronomy, University of Maryland, College Park, MD, 20742-2421, USA
         \and
	 Laboratoire GEOPS (G\'eosciences Paris Sud), Bat. 509, Universit\'e Paris Sud, 91405 Orsay Cedex, France 
         \and
         LESIA, Obs. de Paris, CNRS, Univ Paris 06, Univ. Paris-Diderot, 5 place J. Janssen, 92195 Meudon, France 
         \and
         LATMOS, CNRS/UVSQ/IPSL, 11 boulevard d'Alembert, 78280, Guyancourt, France
         \and
         Centro di Ateneo di Studi ed Attività Spaziali, "Giuseppe Colombo" (CISAS), University of Padova, via Venezia 15, 35131 Padova, Italy
  	 \and
         Department of Industrial Engineering, University of Padova, 35131 Padova, Italy
         \and
         CNR-IFN UOS Padova LUXOR, via Trasea 7, 35131 Padova, Italy
         \and
         UNITN, Universit di Trento, via Mesiano, 77, 38100 Trento, Italy
         \and
         Physikalisches Institut, Sidlerstr. 5, University of Bern, CH-3012 Bern, Switzerland
         \and
         INAF - Osservatorio Astronomico, Via Tiepolo 11, 34143, Trieste, Italy
         \and
         Instituto de Astrofisica de Andaluc\'ia (CSIC), Glorieta de la Astronom\'ia s/n, 18008 Granada, Spain
         \and
         Institute for Space Science, Nat. Central Univ., 300 Chung Da Rd., 32054, Chung-Li, Taiwan
         \and
         Operations Department, European Space Astronomy Centre/ESA, P.O. Box 78, 28691 Villanueva de la Canada, Madrid, Spain
         \and
         Centre for Astrophysics and Planetary Science, Schoold of Physical Sciences (SEPnet), The Univsersity of Kent, Canterbury, CT2 7NH, UK
         \and
         Southwest Research Institute, 1050 Walnut St., Boulder, CO 80302, USA
         \and
         INAF, Osservatorio Astronomico di Padova, 35122 Padova, Italy
         \and
	 University of Padova, Department of Information Engineering, Via Gradenigo 6/B, 35131 Padova, Italy
         \and
         Konkoly Observatory, Budapest H-1525, P.O. Box 67, Hungary
   }
   \date{Received -- ; accepted --}

\titlerunning{Temporal morphological changes in the Imhotep region of comet 67P/Churyumov-Gerasimenko}
\authorrunning{Groussin et al.}

 
  \abstract
   {} 
   {We report on the first major temporal morphological changes observed on the surface of the nucleus of comet 67P/Churyumov-Gerasimenko, in the smooth terrains of the Imhotep region.}
   {We use images of the OSIRIS cameras onboard Rosetta to follow the temporal changes from 24 May 2015 to 11 July 2015.}
   {The morphological changes observed on the surface are visible in the form of roundish features, which are growing in size from a given location in a preferential direction, at a rate of 5.6\,--\,8.1\,$\times$\,10$^{-5}$\,m\,s$^{-1}$ during the observational period. The location where changes started and the contours of the expanding features are bluer than the surroundings, suggesting the presence of ices (H$_2$O and/or CO$_2$) exposed on the surface. However, sublimation of ices alone is not sufficient to explain the observed expanding features. No significant variations in the dust activity pattern are observed during the period of changes.}
   {}

   \keywords{Comets: individual: 67P/Churyumov-Gerasimenko -- Comets: general -- Methods: data analysis}
   \maketitle
%

\section{Introduction}

Comets are among the most primitive bodies of our solar system and contain clues to constrain its formation and evolution \citep[e.g.,][]{Weidenschilling2004}. They are active bodies, which eject gas and dust into space during their orbit around the Sun. A key scientific question, to understand how comets work and whether they still contain pristine materials at or near their surface, is how the nucleus is changing with time and to which extent activity modifies its surface? Rosetta, which has been orbiting comet 67P/Churyumov-Gerasimenko since August 2014, offers a unique opportunity to tackle this fundamental question.

The only changes observed so far at the surface of a comet nucleus are that of 9P/Tempel\,1, visited twice, in 2005 by the Deep Impact spacecraft \citep{AHearn2005} and in 2011 by the Stardust spacecraft \citep{Veverka2013}. The morphological changes are restricted to a small area located near the largest smooth terrain. The two main detected changes are a retreat of up to 50\,m of the boundaries of the smooth flow in at least two places and the merging of three roundish depressions into a larger one \citep{Veverka2013}. 

Turning to the nucleus of 67P, modeling by \cite{Keller2015} predicts that it may locally lose up to 3.5\,--\,14.5\,m per perihelion passage, assuming a dust-to-gas ratio of 4 \citep{Rotundi2015}. The erosion is non-uniform across the surface and strongly connected to insolation, with maximum erosion in the Southern hemisphere, the most illuminated one at perihelion. Observational evidences of past changes also exist on the nucleus surface, in the form of mass wasting \citep{Sierks2015,Thomas2015a} and material transport \citep{Thomas2015b}. Since these changes precede the Rosetta rendezvous, it is not possible to know their timescale and in particular if they result from a single or multiple perihelion passages.

Until 24 May 2015, despite the fact that 67P had already lost 20\% of its total mass loss per perihelion passage \citep{Keller2015}, the OSIRIS cameras \citep{Keller2007} onboard Rosetta had only detected subtle changes on the surface (e.g., in the Hapi region), whose authenticity is still being evaluated. In this article, we report on the first major temporal changes observed on the surface of 67P, in the smooth terrains of the Imhotep region. This region, which presents a wide variety of terrains and morphologies, is a good candidate for being an active region at perihelion \citep{Auger2015,Keller2015,Vincent2013}. The most remarkable features are the smooth terrains, which extend over 0.8\,km$^2$ for the largest one, and the roundish features observed near the gravitational low of the region and interpreted as ancient degassing conduits by \cite{Auger2015}.

\begin{figure*}
\centering
\includegraphics[width=\hsize]{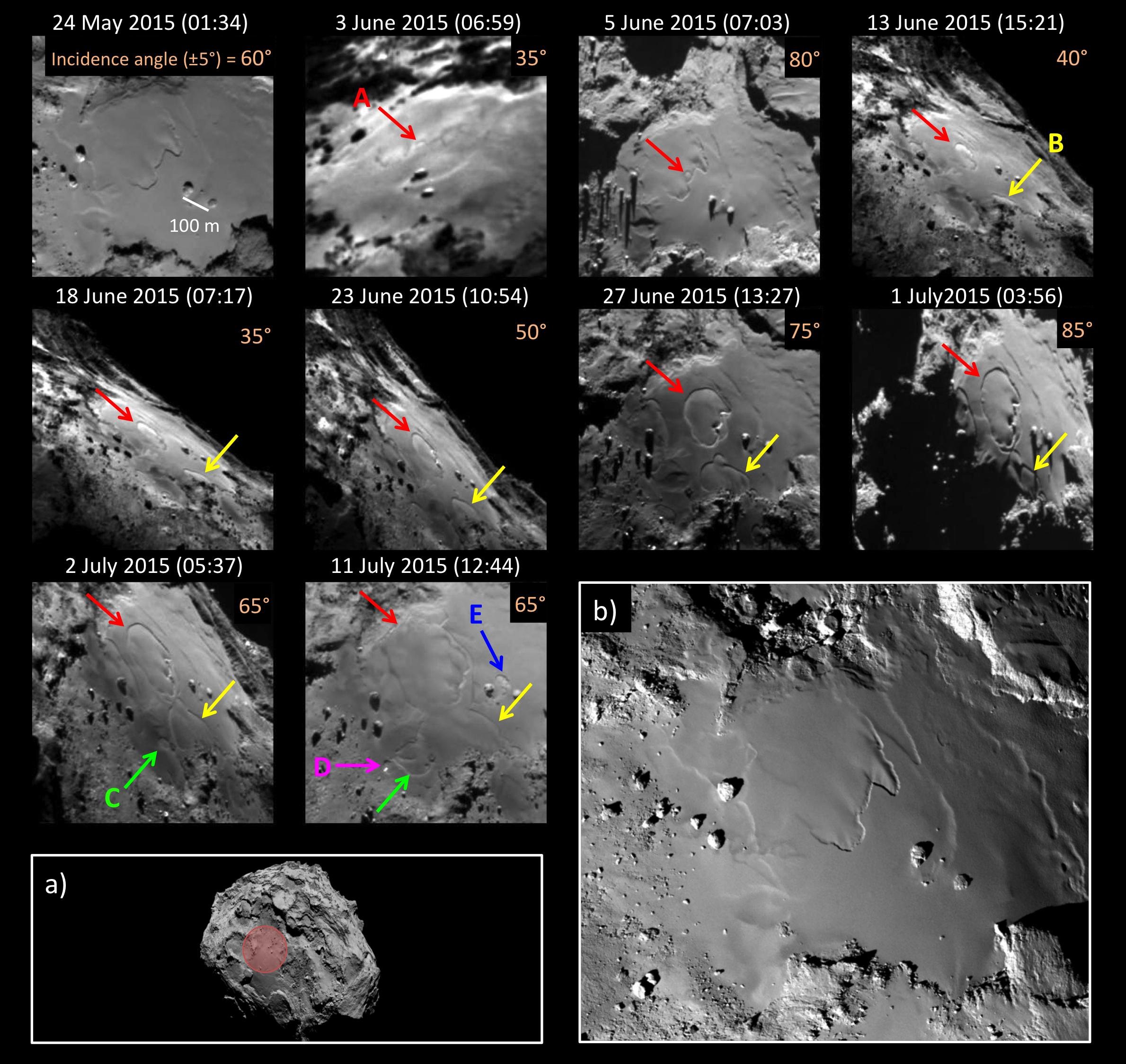}
\caption{Temporal evolution of the smooth terrains of Imhotep, from 24 May 2015 to 11 July 2015 (UT). The arrows indicate the position of the five main detected changes, noted A to E. The spatial resolution improved from 4.3\,m/pix to 3.0\,m/pix over the period. The white line on 24 May shows a 100\,m reference scale, which corresponds to the distance between two remarkable boulders, visible in all images. Panel a) shows the regional context with the region of changes highlighted in red. Panel b) shows the smooth terrains at high spatial resolution (1\,m/pix) in Sept. 2014, before the changes (adapted from \cite{Auger2015}). All the images were acquired with the Narrow Angle Camera (NAC) of the OSIRIS instrument.}
\label{Fig1}
\end{figure*}

\section{Observations}

\subsection{Temporal evolution}

Fig.\,\ref{Fig1} shows the temporal variations of the Imhotep region, from 24 May 2015 to 11 July 2015. Since the Rosetta arrival around 67P in August 2014 until 24 May 2015, no changes had been detected in this region, down to the decimeter scale. Starting 3 June 2015, a first roundish feature (A) appeared on the surface and expanded in the following days. On 13 June 2015, a second roundish feature (B) appeared and also expanded. On 2 July 2015, a third feature (C) appeared and expanded. On 11 July 2015, two additional features (D and E) appeared. At the end of the observational period, features A, B and C joined each others and feature A reached the edges of the smooth terrains.

Each feature is growing in size from a given location in a preferential direction, in a roughly circular pattern. Features originate from scarps (A), cliffs (B), terrain discontinuities (C and E) or edges (D). The expansion rate of the two main features A and B is 8.1\,$\pm$\,0.8\,$\times$\,10$^{-5}$\,m\,s$^{-1}$ and 5.6\,$\pm$\,0.6\,$\times$\,10$^{-5}$\,m\,s$^{-1}$ respectively, this rate being constant from observation to observation. On 2 July 2015, the diameter of feature A reached $\sim$220\,m and $\sim$140\,m for feature B, with rims of 5\,$\pm$\,2\,m. In one month, these dramatic morphological changes have already modified more than 40\% of the surface of the largest smooth terrain of Imhotep, and appear to be still ongoing.

\subsection{Bluish materials appear on the surface}

\begin{figure}
\centering
\includegraphics[width=\hsize]{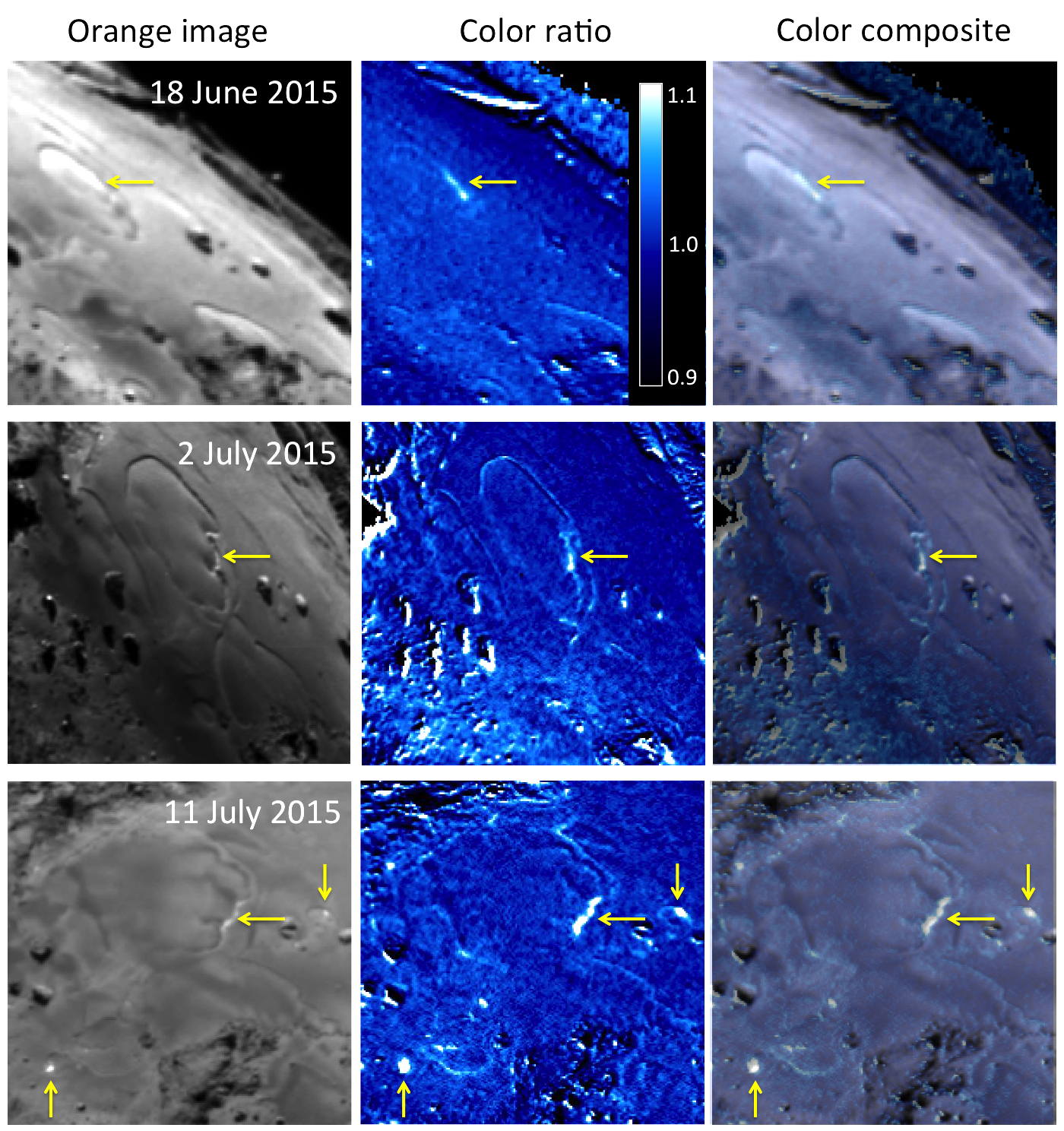}
\caption{Blue (481\,nm) over orange (649\,nm) or red (701\,nm) color ratios of the evolving region on 18 June 2015 (upper row), 2 July 2015 (middle row) and 11 July 2015 (lower row). The first column shows a context image of the region, the second column shows the color ratio (blue/orange for 18 June and 2 July, blue/red for 11 July) and the third column shows a color composite made of the first two columns. The yellow arrows indicate where features A, D and E started (Fig.\,1), with a material bluer than the average (value above 1 in the color ratio). The images were acquired with the NAC.}
\label{Fig2}
\end{figure}

Fig.\,\ref{Fig2} shows the blue/orange or blue/red color ratios of the evolving regions. The terrain where features A, D and E started is bluer than the surroundings, and brighter (best viewed on 11 July). The contours of the expanding features are also slightly bluer. 
The spectrophotometric analysis reveals that the bluest terrains are almost neutral in the range 500\,--\,950\,nm, whereas the spectral slope of the average terrain is 16\,\%\,per\,100\,nm in the same wavelength range, i.e. redder than the Sun.

This bluer material strongly suggests the presence of ice exposed on the surface \citep{Pommerol2015,Fornasier2015,Capaccioni2015}. This fresh material was probably buried below a dust deposit, which has been removed by the erosion processes responsible for the observed changes. The presence of material enriched in volatiles in the first meters below the surface is supported by the nucleus low thermal inertia of 10\,--\,50\,J\,K$^{-1}$\,m$^{-2}$\,s$^{-0.5}$ \citep{Gulkis2015} and modeling \citep[e.g.,][]{Prialnik2004}, which both show that only the top few meters are affected by insolation. Since H$_2$O and CO$_2$ where detected in the coma above Imhotep \citep{Hassig2015}, the ice exposed on the surface could be composed of H$_2$O and/or CO$_2$.

\subsection{Link with dust activity}

\begin{figure}
\centering
\includegraphics[width=\hsize]{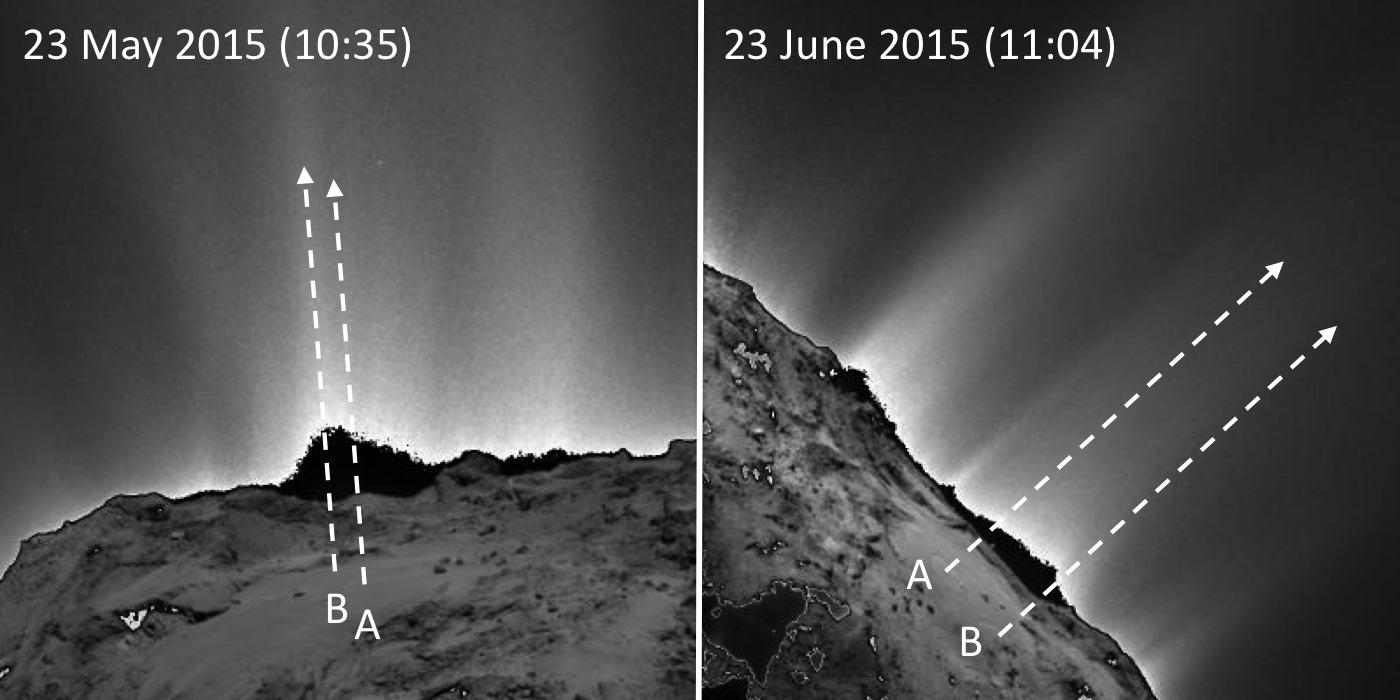}
\caption{Activity above Imhotep on 23 May 2015 and 23 June 2015 (UT), before and after the changes started. The nucleus local time differs by less than 30\,min between the 2 images.
The white arrows indicate the position where we would expect an increase in activity, if jets were coming from features A or B. The images were acquired with the NAC, more than 50$^{\circ}$ away from nadir.}
\label{Fig4}
\end{figure}

We monitored the dust activity in the Imhotep region before and during the period of changes (Fig.\,\ref{Fig4}). We did not detect any significant variations in the dust activity pattern above this region and in particular no increase in the number and intensity of dust jets above features A and B. More precisely, whereas several narrow collimated jets might originate from the smooth region, no strong jets are associated specifically with features A or B. 

Although no significant variations are observed in the dust activity, we cannot exclude an increase in the gas production rate (H$_2$0, CO$_2$ or CO) during the observational period, unfortunately beyond the capabilities of the OSIRIS instrument. As an example, on comet 103P/Hartley\,2, the water and dust strongest jets were not originating from the same region; the neck for water and the small lobe for the dust \citep{AHearn2011}.


\section{Discussions and conclusions}

We observe a collapse of a considerable thickness ($\sim$5\,m) of the upper surface, which is occurring in an organized way and propagating over the surface. This process starts from scarps, cliffs, terrain discontinuities or edges, where the dust deposit is thinner and where the seasonal heat wave first reaches the underneath volatile rich materials. In May\,--\,July 2015, these terrains are orientated towards the afternoon side; they are strongly illuminated in the morning and have the Sun at zenith above them, so they are probably local spots hotter than the neighbourhood. The expansion goes preferentially away from these scarps or cliffs, which are consolidated materials, harder to erode. Later on, it may overpass this natural barrier, as it is the case for feature A on 27 June (Fig.\,1). Each feature continues to grow in size, towards the afternoon direction, until it reaches another one or the edges of the smooth terrain, and then stops.

We calculated that the erosion rate for H$_2$O and CO$_2$ ice exposed on the surface can reach 4.2\,$\times$\,10$^{-7}$\,m\,s$^{-1}$ and 2.8\,$\times$\,10$^{-6}$\,m\,s$^{-1}$ respectively at 1.4\,au (67P's heliocentric distance in June 2015), assuming a density of 535\,kg\,m$^{-3}$ \citep{Preusker2015}. This is much less than the observed lateral expansion rate of 5.6\,--8.1\,$\times$\,10$^{-5}$\,m\,s$^{-1}$, which means that the sublimation of ices alone is not sufficient to explain the observed expanding features. Although it remains speculative at this stage, the eroding process might however be exacerbated by the low tensile strength of tens of Pa of the cometary material \citep{Groussin2015} that can easily erode in large chunks of several meters \citep{Pajola2015} or driven by additional energetic processes such as for example clathrate destabilization or amorphous water ice crystallization \citep{Mousis2015}.

The lack of enhanced dust activity in the coma during the period of changes suggests that the dust deposit is dominated by millimetric or larger particles, and not microscopic ones. Grains up to the decimeter scale are observed in the smooth terrains of Imhotep \citep{Auger2015}. During the eroding process, a fraction of dust particles fall and accumulate at the foot of the expanding features or are transported slightly farther away, towards the center of the newly formed roundish features. This material transport could explain the bumpy shape of feature A (1 July, Fig\,1), with more material in the center than on the edges. The fraction of particles that escape the nucleus is unkown and cannot be determined by OSIRIS alone. We further add that if the nucleus is porous, a significant fraction of dust particles may sink into it during this process. Finally, it is worth mentioning that the surface material in the evolving regions looks photometrically essentially the same, before and after the changes (Fig.\,1).



If the same erosion process applies to the putative ancient degassing conduits of Imhotep \citep{Auger2015}, covered by a dust deposit for most of them, they could also be rejuvenated and become active at each perihelion passage. More generally, if activity increases significantly on Imhotep, it could be the source of the strong equatorial jet observed from Earth at the previous perihelion passage \citep{Vincent2013}. Such seasonal events are also supported by the fact that the height of the rims of the expanding features is a few meters, which corresponds to the depth at which the seasonal heat wave penetrates into the nucleus \citep{Gulkis2015}.

One may wonder why such major and rapid changes on the surface are first and currently only observed in the smooth terrains of Imhotep. The Northern hemisphere, which is also covered in several places by a dust deposit \citep{ElMaarry2015a}, receives ten times less solar energy than Imhotep and is colder on average \citep{Keller2015}; if the same process takes place (e.g., in the Hapi region), it is on a longer timescale. The Southern hemisphere is, on average, strongly eroded by several meters at each perihelion passage and therefore lacks dust deposits. Imhotep, which is located in a gravitational low able to retain and accumulate dust deposits, and where solar energy reaches 70--80\% of the maximum energy received by the nucleus \citep{Keller2015}, is the proper location for such changes to take place. 

It is therefore tempting to see similarities with the morphological changes observed on 9P between 2005 and 2011 (Sect.\,1), which are also linked to the largest smooth terrain on the surface, furthermore in a gravitational low like Imhotep \citep{Thomas2007}. The retreat of the scarp from the smooth flow is best interpreted by the progressive sublimation of material over several months \citep{Veverka2013}, i.e. a slow process compared to what we see on Imhotep. However, the merging of multiple depressions could be another example of the type of fast eroding process observed on Imhotep, with the strong limitation that we do not know the timescale of this event on 9P.

The dramatic changes observed on Imhotep are a spectacular event, unique to comets, with a currently unpredictable endstate. We will continue to carefully monitor this region during the coming months to better constrain the erosion processes responsible for these changes. Looking for changes on 67P remains a key scientific objective for all Rosetta instruments, to better understand how comets work and evolve.

\begin{acknowledgements}
OSIRIS was built by a consortium of the Max-Planck-Institut für Sonnensystemforschung, Katlenburg-Lindau, Germany; CISAS University of Padova, Italy; the Laboratoire d'Astrophysique de Marseille, France; the Instituto de Astrofísica de Andalucia, CSIC, Granada, Spain; the Research and Scientific Support Department of the ESA, Noordwijk, Netherlands; the Instituto Nacional de Técnica Aeroespacial, Madrid, Spain; the Universidad Politéchnica de Madrid, Spain; the Department of Physics and Astronomy of Uppsala University, Sweden; and the Institut für Datentechnik und Kommunikationsnetze der Technischen Universität Braunschweig, Germany. The support of the national funding agencies of Germany (DLR), France (CNES), Italy (ASI), Spain (MEC), Sweden (SNSB), and the ESA Technical Directorate is gratefully acknowledged. We thank the Rosetta Science Operations Centre and the Rosetta Mission Operations Centre for the successful rendezvous with comet 67P/Churyumov-Gerasimenko.
\end{acknowledgements}


\bibliographystyle{aa}
\bibliography{References_Imhotep}


\end{document}